%% file: main.tex
\documentclass[runningheads]{llncs}

\usepackage{graphicx}
\usepackage{subfigure}
\usepackage{amsmath}
\usepackage{hyperref}

\begin{document}

\title{iDataCool: HPC with Hot-Water Cooling and Energy Reuse}

\author{%
Nils Meyer\inst{1} \and 
Manfred Ries\inst{2} \and 
Stefan Solbrig\inst{1} \and 
Tilo Wettig\inst{1}
}

\institute{%
Department of Physics, University of Regensburg, 93040 Regensburg, Germany \and
IBM Deutschland Research \& Development GmbH, 71032 B\"oblingen, Germany
}

\maketitle

\setcounter{footnote}{0}

\begin{abstract}
\input{abstract}
\end{abstract}

\section{Introduction}
\input{introduction}

\section{iDataCool architecture}
\label{sec:idc-arch}
\input{idc-arch}

\section{Infrastructure}
\label{sec:installation}
\input{installation}

\section{Measurements}
\label{sec:benchmarks}
\input{benchmarks}

\section{Conclusions}
\input{conclusions}

\section*{Acknowledgments}
\input{acknowledgment}

\bibliographystyle{splncs}

\end{document}

%% file: abstract.tex
iDataCool is an HPC architecture jointly developed by the University
of Regensburg and the IBM Research and Development Lab B\"oblingen.
It is based on IBM's iDataPlex platform, whose air-cooling solution
was replaced by a custom water-cooling solution that allows for
cooling water temperatures of 70$^\circ$C/158$^\circ$F.  The system is
coupled to an adsorption chiller by InvenSor that operates efficiently
at these temperatures.  Thus a significant portion of the energy spent
on HPC can be recovered in the form of chilled water, which can then
be used to cool other parts of the computing center.  We describe the
architecture of iDataCool and present benchmarks of the cooling
performance and the energy (reuse) efficiency.


%% file: introduction.tex
According to a 2012 IDC study \cite{IDC}, the worldwide costs for
power and cooling of IT equipment now exceed 25 billion US-\$ per year
and are comparable with the costs for new hardware.  For this obvious
financial reason, but also because of the impact on the environment,
energy efficiency has become a very important concern in the IT
industry.  The problem can be addressed in two ways.  First, every
effort should be made to reduce the energy consumed by the equipment.
Second, some of the energy could be reused.  In this paper we will
address both of these points, concentrating on the cooling part in
``power and cooling''.  We present an innovative liquid-cooling
solution for a high-performance computing (HPC) system that allows for
free cooling year-round and energy reuse in the form of chilled-water
generation.

A discussion of various aspects of liquid cooling with focus on high
coolant temperatures can be found in Refs.~\cite{hot,deep7.1}.  For
the following discussion, we assume that the cooling medium is water
and define what we mean by ``warm water'' and ``hot water''.  We
consider water to be warm if its temperature is higher than the
wet-bulb temperature of the ambient air even on hot days so that free
cooling is always possible.  In typical climates this means about
40$^\circ$C/104$^\circ$F.  Free cooling year-round drastically reduces
the cooling costs since chillers are no longer needed.  Even some
possibilities for energy reuse exist, e.g., the warm water could drive
an underfloor heating system.  We consider water to be hot if it opens
up more possibilities for energy reuse, e.g., if it is hot enough to
drive a radiator-based heating system or an adsorption chiller.  This
means at least 65$^\circ$C/149$^\circ$F.  Sustaining such
cooling-water temperatures in a large system that is running in stable
production mode over long periods of time is a real problem, which we
claim to have solved in the project iDataCool described in this paper.
The main innovation of iDataCool is the design of a low-cost processor
heat sink that minimizes the temperature difference between cooling
water and processor and allows for cooling-water temperatures of up to
70$^\circ$C/158$^\circ$F.

In the project described in this paper, the infrastructure conditions
at the installation site are such that reusing energy for heating
purposes is not an option.  Therefore the hot water was used to drive
an adsorption chiller that generates chilled water.  This is another
innovation of iDataCool, which is of potential interest for computing
centers in hot climates.

Related projects with similar goals (i.e., hot-water cooling) are
Aquasar \cite{aquasar} and CoolMUC \cite{coolmuc}.  Both of these are
somewhat smaller in scale and run at somewhat lower temperatures.
There are also a number of projects that allow for warm-water cooling
as defined above, e.g., \cite{qpace,aurora,csc,supermuc}.


%% file: idc-arch.tex
\begin{figure}[t]
  \begin{center}
    \centerline{\includegraphics[width=0.47\textwidth]{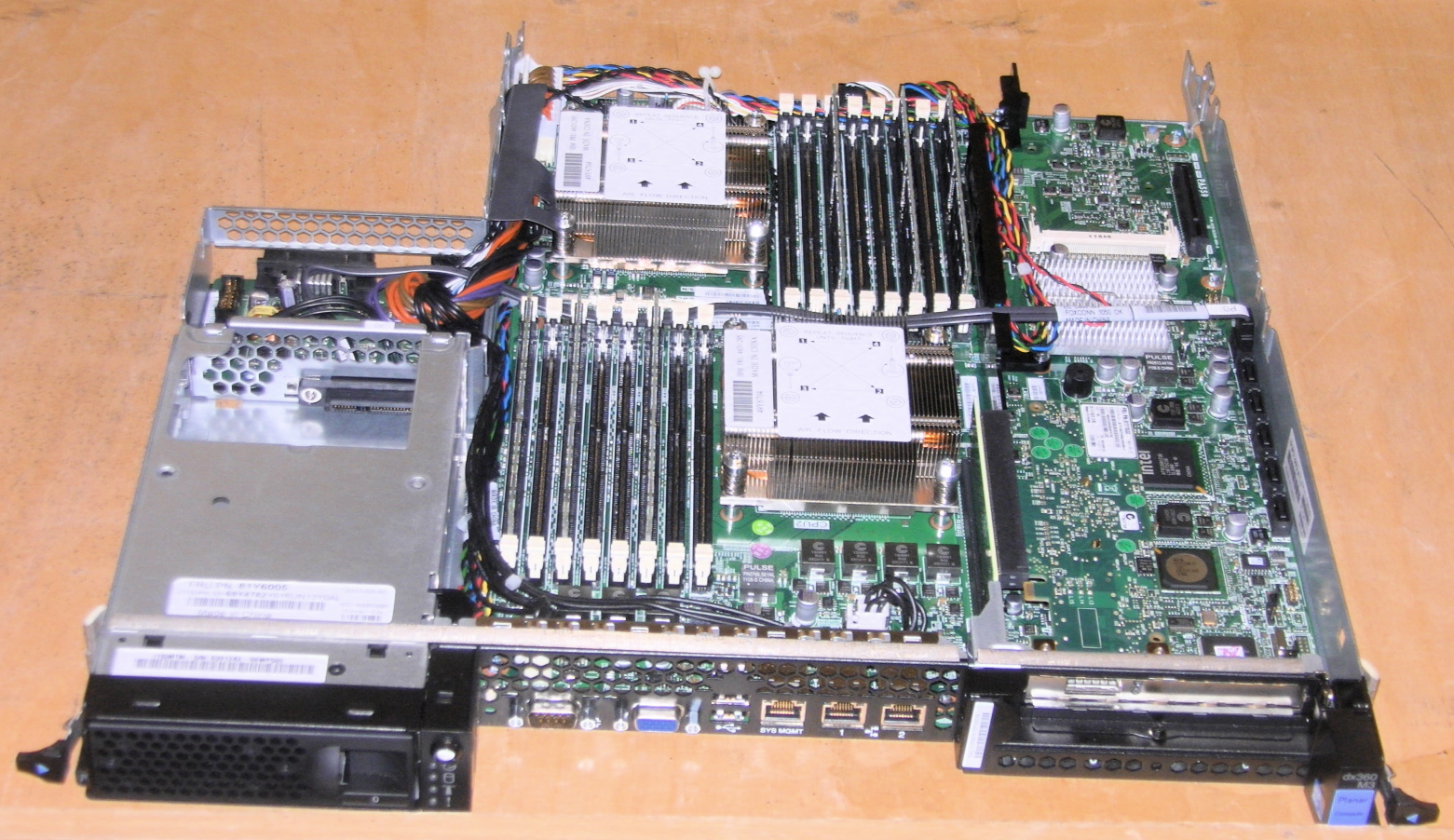}\hfill%
      \includegraphics[width=0.47\textwidth]{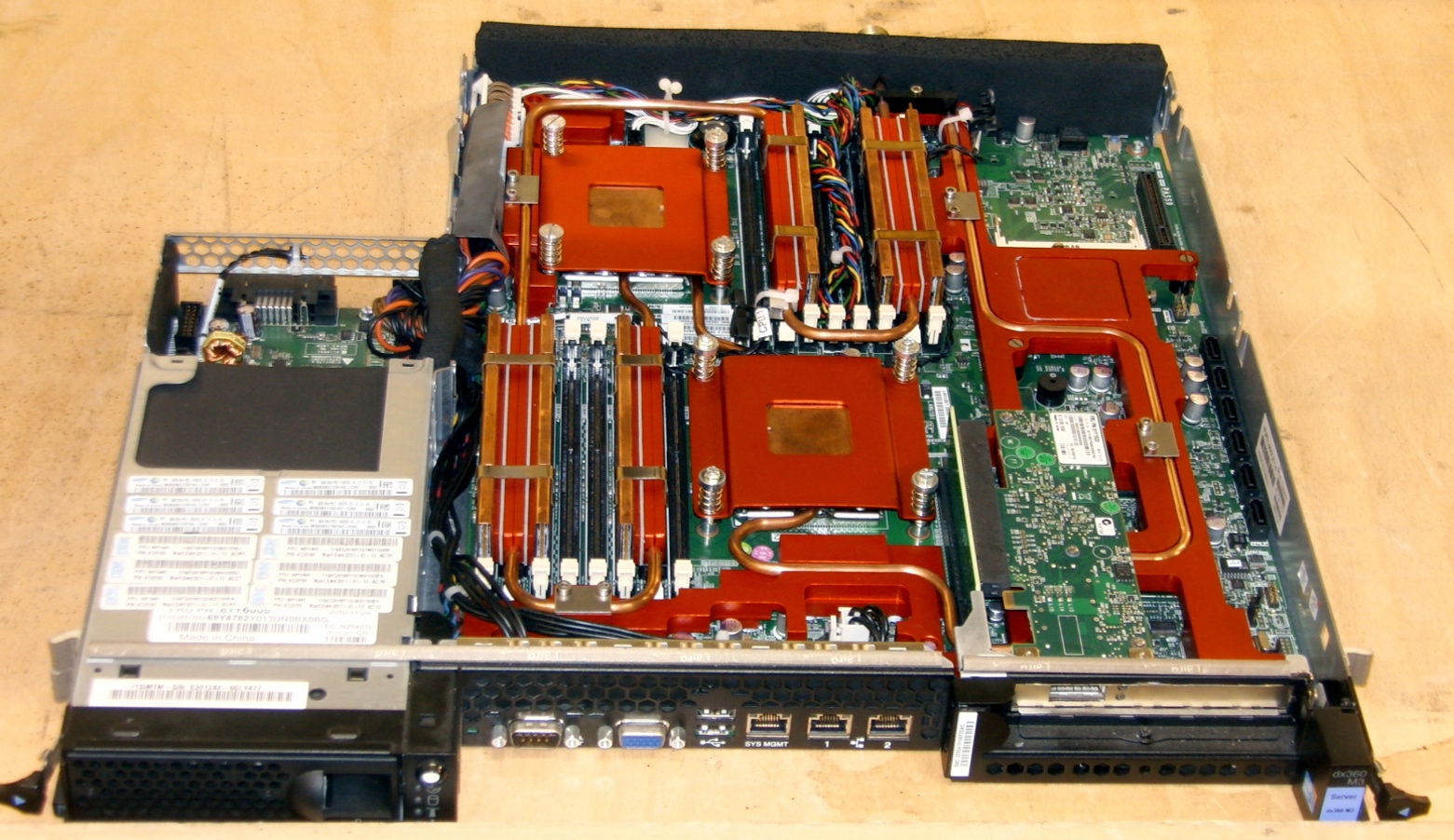}}
    \caption{Left: Original air-cooled iDataPlex dx360 M3 compute
      node. The power supply unit (not shown) is attached to the
      compute node on the top left. Right: iDataCool compute node with
      new water-cooling solution, consisting of a copper pipeline,
      copper heat sinks for processors and memory, and aluminium heat
      bridges.  Armaflex thermal insulation is used to prevent heat
      from escaping into the environment.}
    \label{fig:idc-node}
  \end{center}
\end{figure}

Before presenting our liquid-cooling solution we briefly describe the
iDataCool HPC cluster, which is based on the IBM System x iDataPlex
architecture \cite{idataplex}.  It consists of three racks with 72
compute nodes each.  A compute node is equipped with either two
four-core Intel Xeon E5630 (44 in total) or two six-core Intel Xeon
E5645 (388 in total) Westmere processors organized as a distributed
shared memory system.  Each node contains 24 GB of memory arranged in
six 4 GB DDR3 dual in-line memory modules. The main interconnect
network of iDataCool is based on QDR Infiniband, arranged in a hybrid
ring/tree topology. Switched Gigabit Ethernet is used for disk I/O,
system booting via NFS, and job scheduling. Every compute node is
monitored and controlled by a dedicated baseboard management
controller (BMC).

\begin{figure}[t]
  \centering
  \includegraphics[scale=.7]{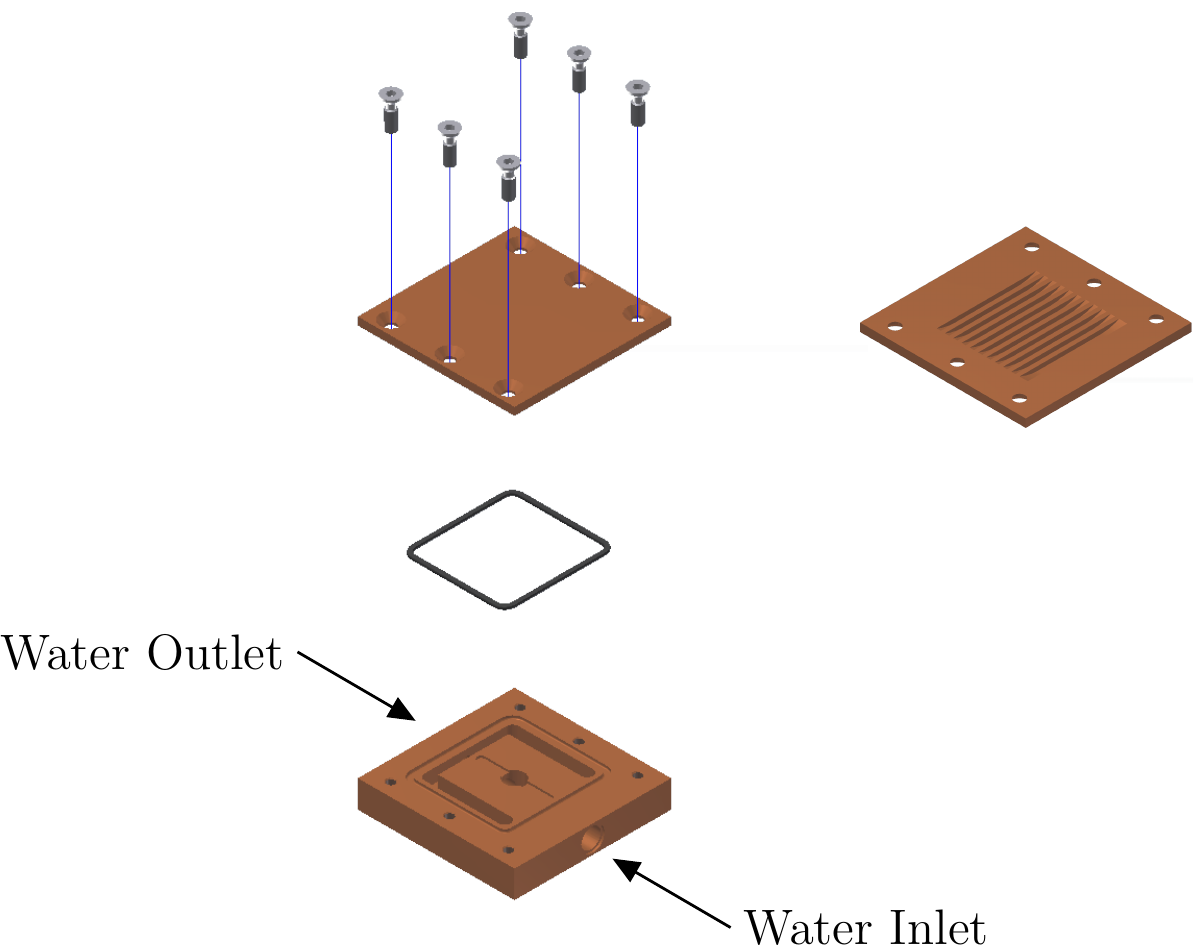}
  \caption{Design of the iDataCool heat sink.  The top part, which is
    attached to the processor package, is shown from both sides.}
  \label{fig:heatsink}
\end{figure}

The air-cooling components of the original iDataPlex system were
completely removed and replaced by a custom water-cooling solution,
shown in Fig.~\ref{fig:idc-node}, which was developed in a joint
effort of the University of Regensburg and the IBM Research and
Development Lab B\"oblingen and manufactured and installed in the
machine shop of the Regensburg Physics Department.  The main design
drivers of the water-cooling solution were the possibility of
hot-water cooling and low cost.  Let us focus on the first point for
now.  CPUs can tolerate a certain maximum chip temperature, which
depends on the specific chip used.  Thus, to enable high cooling-water
temperatures, the temperature difference between the compute cores and
the water should be minimized.  The heat transfer path can be divided
into two segments.  First, the heat is transferred from the cores to
the package surface.  Second, the heat is transferred from the package
surface to the cooling water via thermal interface material and a heat
sink.  We have no control over the first segment but can optimize the
second one, in particular through the design of the processor heat
sink, which is shown in Fig.~\ref{fig:heatsink}.  Its design
parameters were as follows.

\begin{itemize}
\item Minimize the temperature difference between coolant and
  processor package.  This was achieved by bringing the coolant very
  close to the package and by using a material with high thermal
  conductivity, i.e., copper. (For the thermal interface material
  between heat sink and processor package we used Shin-Etsu
  X23-7783D.)
\item Efficient thermal transport.  This was achieved by the design
  shown in Fig.~\ref{fig:heatsink}, which provides a sufficiently
  large interface area for heat transfer and creates turbulent flow.
\item Low pressure drop.  The channels shown in
  Fig.~\ref{fig:heatsink} are not microchannels but 1 mm wide.  At a
  typical flow rate of 0.6 l/min the pressure drop is less than 0.1
  bar.
\item Low cost.  The design shown in Fig.~\ref{fig:heatsink} is very
  simple and can be manufactured inexpensively with standard tools
  since the channels are rather wide.  Using an O-ring and screws is
  simpler and more leak-proof than using glue.
\end{itemize}

The processor heat sinks are hard-soldered to a copper pipeline that
provides the water flow.  Other heat-critical components on the board
are thermally coupled to the pipeline via copper or aluminum heat
bridges and thermal interface material.  These components include
memory modules, Infiniband daughter card, chip set, voltage
regulators, and several other chips.  All of these components can
tolerate higher temperatures than the processors.  Different materials
and designs are used to satisfy the cooling needs of these parts.
E.g., the memory modules are cooled via copper heat bridges clamped to
aluminum bars which embrace the cooling pipeline.  Thus the memory
modules can easily be replaced in the field.  Proper mounting
(including thermal interface material) and alignment of all components
of the cooling solution is crucial for a high cooling performance.

Heat dissipation into the environment of the compute node is reduced
using Armaflex thermal insulation.  The only components of iDataCool
that are still air-cooled are the power supply units and the network
switches.

Only a minor modification of the node chassis was required to connect
the cooling pipeline, via inexpensive standard water connectors, to a
rack-level manifold.  The nodes are connected to the manifold in a
parallel fashion.  The manifold is designed using the Tichelmann
principle to ensure that the distance covered by the water flow, and
therefore the pressure drop, is equal for all nodes.  Thus the water
flow rates balance themselves automatically.  The manifold is attached
to the backside of each rack. Armaflex thermal insulation reduces the
dissipation of heat from the pipes into the computing center.

An important issue is the cost of the liquid-cooling solution.  All
components are made from standard materials (copper, aluminum,
plastic) and were designed such that the manufacturing process is
simple, and thus inexpensive.  There are only six soldering joints per
node (two at each heat sink, and one each at the in- and outlet), and
the bending of the copper pipe can be automated by a properly designed
tool.  The mounting of the liquid-cooling solution (including
application of thermal interface material) was somewhat
time-consuming, but on an industrial scale this process could also be
automated.  For us the total cost of the liquid-cooling solution was
about 120 Euro per node (excluding external infrastructure).  While
this is more expensive than an air-cooled solution, it is a small
fraction of the overall cost and can be amortized quickly by the
savings from free cooling and energy reuse.  On an industrial
scale the costs would probably be even lower.

Our sensing and monitoring facilities are described at the beginning
of Sect.~\ref{sec:benchmarks}.


%% file: installation.tex
\begin{figure}[t]
  \centerline{\includegraphics[width=\textwidth]{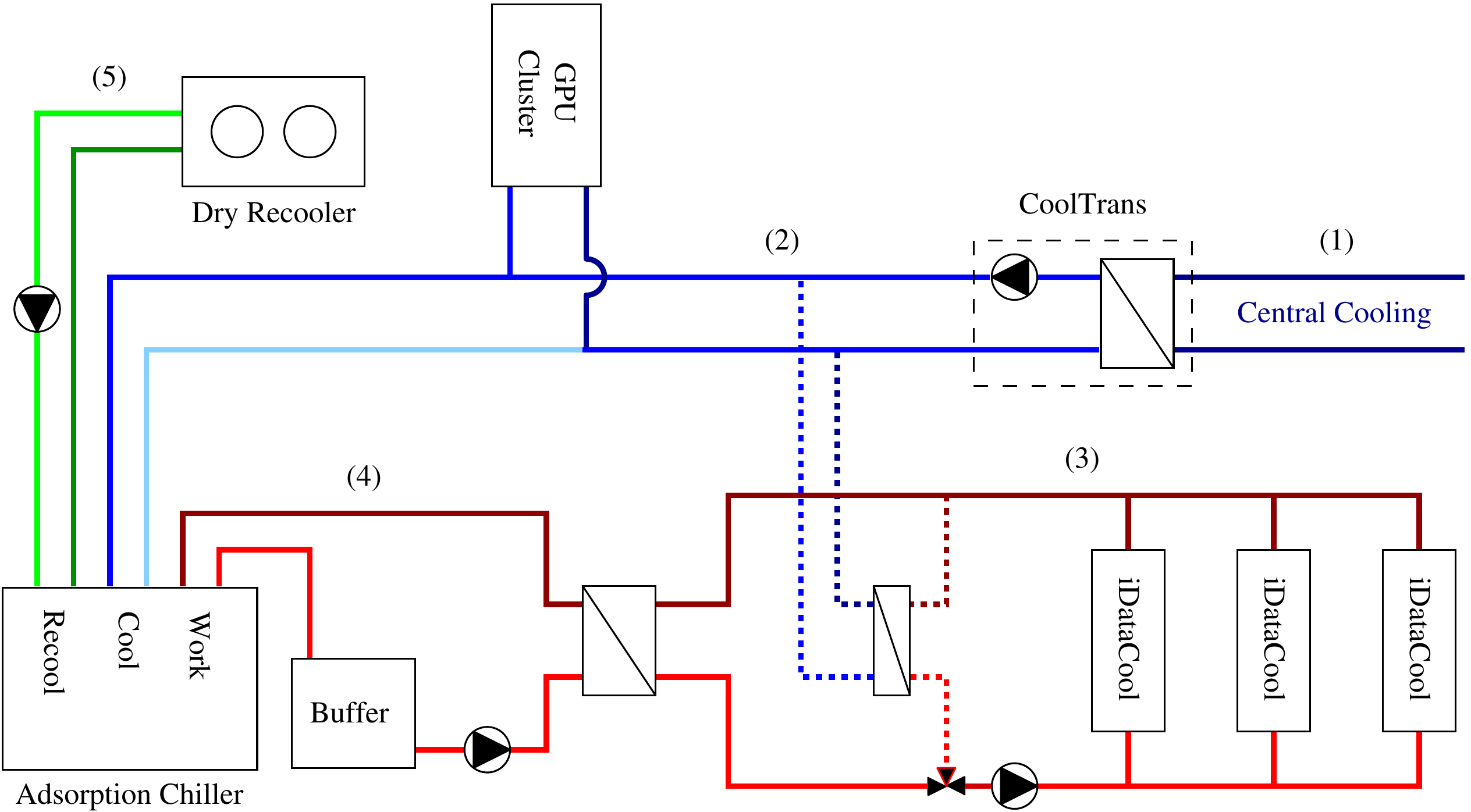}}
  \caption{Liquid-cooling installation consisting of central cooling
    circuit~(1), primary cooling circuit~(2), rack cooling
    circuit~(3), driving circuit~(4), and recooling circuit~(5).}
\label{fig:circuits}
\end{figure}

iDataCool is installed in the computing center of Regensburg
University, which was entirely air-cooled before.  The liquid-cooling
infrastructure for iDataCool was prepared in 2011 and completed in
2012.  Since the spring of 2012 the waste heat of iDataCool drives an
adsorption chiller (LTC 09 by InvenSor), which in turn generates
chilled water.  The LTC 09 is a so-called low-temperature chiller that
works efficiently already at driving temperatures of around
65$^\circ$C/149$^\circ$F, see the efficiency curves in the data sheet
\cite{invensor}.  The cooling performance of the chiller is balanced
against the cooling needs of a small-sized GPU cluster that is cooled
by the LTC 09.  The GPU cluster has a peak power consumption of 12kW.
The equipment of the GPU cluster is housed in a closed cabinet and
cooled by air. The air is cooled by an air-water heat exchanger
(Kn\"urr CoolLoop \cite{coolloop}) that transfers the heat inside the
cabinet to the water circuit.

Fig.~\ref{fig:circuits} shows a schematic overview of the
liquid-cooling installation.  It consists of five water circuits.
Each circuit is driven by a dedicated pump that keeps the water flow
at a constant rate.  Energy losses to the environment are reduced by
thermal insulation of the hot parts of the plumbing. Special
additives are used to minimize the risk of corrosion.  In the
following we discuss the details of the five circuits.
\begin{itemize}
\item The computing center is connected to the university's central
  cooling circuit~(1) which delivers chilled water at temperatures
  around 8$^\circ$C/46$^\circ$F.
\item The primary cooling circuit (2) is continuously chilled by the
  adsorption chiller and picks up heat from the GPU cluster.  In addition,
  the primary circuit can be used as an additional cooler for the
  iDataCool cluster, see the dotted lines in the figure. (A dry
  recooler would also suit this purpose.)  If the water
  temperature exceeds 20$^\circ$C/68$^\circ$F the primary circuit is
  supported by the central cooling circuit (1), to which it is
  connected via a commercial heat exchanger that works autonomously
  (Kn\"urr CoolTrans \cite{cooltrans}).
\item The iDataCool cluster is cooled by the rack cooling circuit (3)
  with hot water at outlet temperatures of up to
  $70^\circ$C/158$^\circ$F.  The waste heat of iDataCool is supplied
  to the driving circuit of the adsorption chiller (4) via a heat
  exchanger.  Transfer of excess heat to the primary cooling circuit
  (2) allows us to keep the rack inlet temperature constant (even
  under change of load on the cluster).  The heat transfer to primary
  and driving circuit is continuously regulated by a 3-way valve. The
  valve is automatically operated by a PID controller that determines
  the rack inlet temperature.
\item The adsorption chiller is driven by the driving circuit (4).
  Temperature fluctuations in the driving circuit due to the
  operational characteristics of the chiller are smoothed by a buffer
  tank with a capacity of 800 liters.  Due to proper thermal
  insulation there is virtually no temperature loss at the interface
  to the rack cooling circuit.
\item The recooling circuit (5) connects to a fan-driven dry recooler
  that is located outside the computing center. The fans are
  controlled automatically by the adsorption chiller with the fan
  speed optimized for energy-efficient operation of the chiller.
  Evaporative cooling is possible in principle but has not been
  implemented in our setup.  Freezing of the external recooling
  circuit is avoided by an admixture of ethylene
  glycol.\footnote{Since driving and recooling circuit are connected
    in the chiller, we also have glycol in the driving circuit.  This
    is the reason for the heat exchanger between rack and driving
    circuit.}
\end{itemize}

The standard use case of the adsorption chiller is rather different
from our setup.  Normally the chiller drives an air-conditioning
system, i.e., one specifies the desired temperature of the chilled
water, and the chiller then absorbs as much heat as necessary from the
driving circuit to deliver the required cooling power.  In our case we
want the chiller to absorb the heat from the rack circuit and to
deliver as much cooling power as possible.  To see how this works we
now discuss in some detail the behavior of our system.  The chiller is
characterized by its cooling capacity $P_c^\text{max}$, which is the
maximum amount of heat per unit time it can remove from the cooling
circuit, and by its coefficient of performance, defined as 
\begin{equation}
  \text{COP} = \frac{\text{power}\;P_c\;\text{removed from cooling circuit}}%
  {\text{power}\;P_d^\text{abs}\;\text{absorbed from driving circuit}}\;.\nonumber
\end{equation}
All of these quantities depend (among other parameters) on the
temperature $T$ in the driving circuit \cite{invensor}.  Now assume
that the 3-way valve in Fig.~\ref{fig:circuits} completely shuts off
the additional cooling path and that we turn on the iDataCool cluster
with an initial water temperature of, say, 20$^\circ$C/68$^\circ$F.
At $T<55^\circ$C/131$^\circ$F the adsorption chiller is in standby
mode and thus absorbs no heat from the cluster.  As a result, the
temperature in the rack circuit increases until it goes above
55$^\circ$C/131$^\circ$F and the chiller turns on.\footnote{In our
  system the thermal contact between rack circuit and driving circuit
  is very good so that the driving temperature $T$ equals the outlet
  temperature of the rack.}  What happens then depends on the
temperature dependence of the function
$P_d^\text{max}(T)=P_c^\text{max}(T)/\text{COP}(T)$, which is the
maximum power that can be removed from the driving circuit of the
chiller.  This function depends on the parameters of the chiller.  A
certain power $P_d$ is transferred from the rack circuit to the
driving circuit.  If $P_d^\text{max}(T)<P_d$ the temperature keeps
going up.  If $P_d^\text{max}(T)$ intersects $P_d$ at some
$T=T_\text{eq}$, the system settles into equilibrium at that
temperature.  If $P_d$ is larger than the maximum of
$P_d^\text{max}(T)$, we have to employ the additional cooling
mechanism via the 3-way valve to remove the rest of the heat and to
keep the rack circuit at a well-defined temperature.  The parameters
of our system are such that for
$T=60\ldots70^\circ$C/$140\ldots158^\circ$F the value of
$P_d^\text{max}(T)$ is almost equal to, but slightly smaller than, the
power transferred from the rack circuit to the driving circuit at
maximum load of the cluster. Thus the system is almost in equilibrium
and only a very small amount of additional cooling is necessary.

Our setup also solves two redundancy issues: (i) Should the adsorption
chiller fail to absorb all the heat from the iDataCool cluster,
additional cooling is provided by the primary cooling circuit, which
may be supported by the central cooling circuit. 
(ii) Should the adsorption chiller fail to provide enough
cooling power to the GPU cluster, again the central cooling circuit
comes to the rescue.  


%% file: benchmarks.tex
In this section we present a number of measurements and benchmarks
performed on the iDataCool system.  We first describe our sensing and
monitoring facilities.  The liquid-cooling installation is constantly
monitored, and relevant system parameters are logged electronically.
On the node level, we read out the individual processor core
temperatures from chip-internal sensors, we estimate the water in- and
outlet temperature of each node using the original air-flow
temperature sensors (which we attached to the copper pipe), and we
monitor the DC power consumption of each node.  On the cluster level,
we measure the in- and outlet temperature and the AC power consumption
of the 3-rack installation.  Our instrumentation also allows us to
determine the combined AC power consumption of the iDataCool cluster,
the GPU cluster, water pumps, the adsorption chiller, and the dry
recooler.  To determine the flow rates in the different water circuits
we use various kinds of flow meters.  As for the accuracy of our
equipment, we estimate the node-level temperature sensors to be
accurate to about 1$^\circ$C/2$^\circ$F, while the cluster-level
temperature sensors, which are in direct contact with the water, are
specified to have an accuracy of 0.2$^\circ$C/0.4$^\circ$F.  The
ultrasonic flow meter for the rack cooling circuit is specified to
have an accuracy of 1\%, while the flow meters for the other circuits
are much simpler and only about 10\% accurate. 

When plotting quantities as a function of the cooling-water
temperature we have the choice of using the rack inlet or outlet
temperature.  We chose the outlet temperature $T_\text{out}$ since
this is the quantity of interest for energy reuse purposes.  The
difference between inlet and outlet temperature can be controlled by
adjusting the water flow rate and is about 5$^\circ$C/9$^\circ$F in
our system.\footnote{At constant water flow rate this temperature
  difference decreases somewhat with the outlet temperature since the
  system is not perfectly insulated from the environment.  At higher
  temperatures more heat is lost to the air.}  For constant rack inlet
temperature, $T_\text{out}$ fluctuates slightly depending on the load
of the cluster and on the control parameters.  In the figures, the
horizontal error bars in the $T_\text{out}$ direction reflect these
fluctuations in time.

\begin{figure}[t]
  \centering
  \subfigure[][\label{fig:core}]%
  {\includegraphics{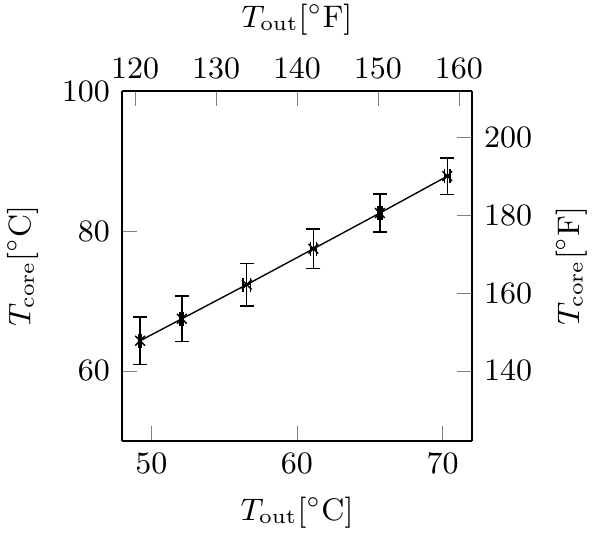}}\hfill
  \subfigure[][\label{fig:coredist}]%
  {\includegraphics{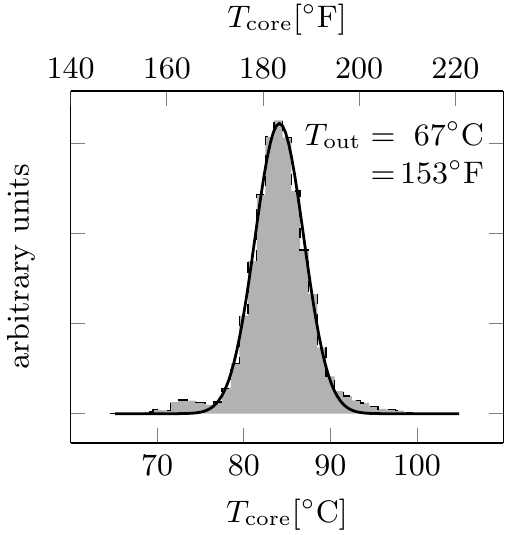}}\\
  \vspace*{-3mm}
  \caption{\subref{fig:core} Core temperatures of 13 compute nodes
    under \texttt{stress} and \subref{fig:coredist} core temperature
    distribution of the cluster in production mode.  In both plots
    only six-core E5645 processors are included.  The vertical error
    bars in \subref{fig:core} are the standard deviations after
    averaging over time and nodes.}
\end{figure}

\begin{figure}[t]
  \centering
  \subfigure[][\label{fig:power}]%
  {\includegraphics{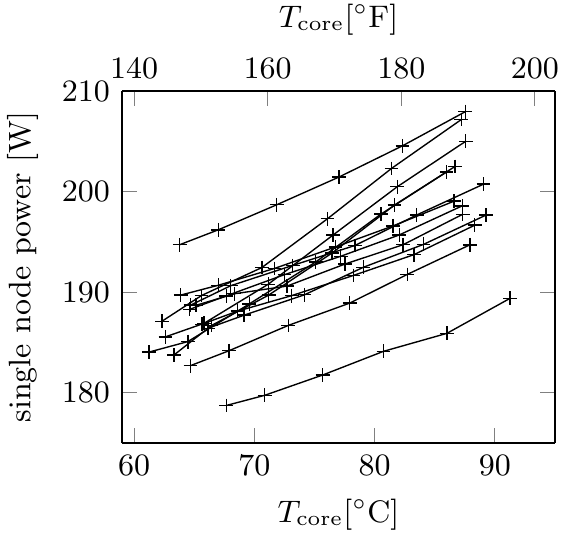}}\hfill
  \subfigure[][\label{fig:powerdist}]%
  {\includegraphics{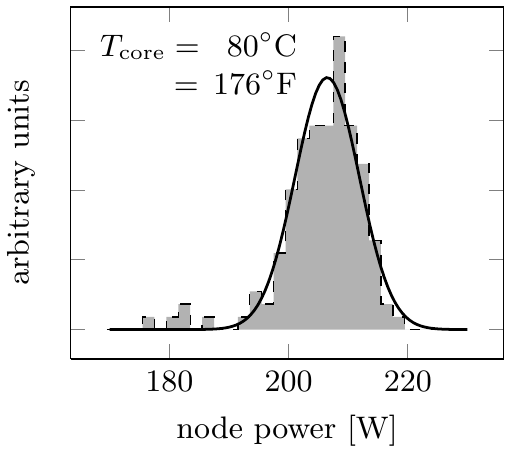}}
  \vspace*{-3mm}
  \caption{\subref{fig:power} Node power consumption of 13 compute
    nodes under \texttt{stress} and \subref{fig:powerdist} node power
    distribution of the cluster in production mode.  In both plots
    only six-core E5645 processors are included. }
\end{figure}

\begin{figure}
  \centering
  \subfigure[][\label{fig:increase}]%
  {\includegraphics{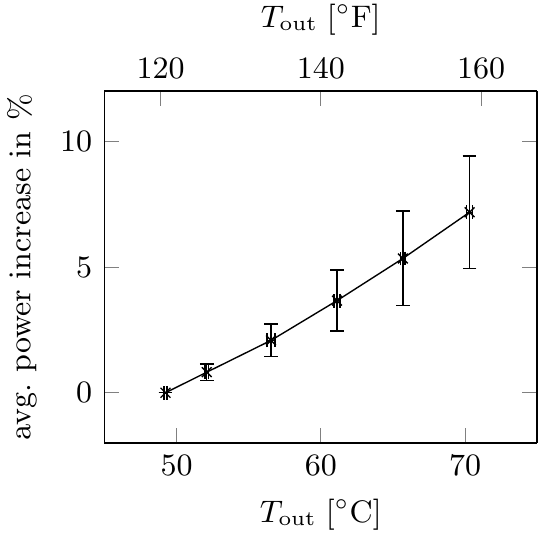}}\hfill
  \subfigure[][\label{fig:cop}]%
  {\includegraphics{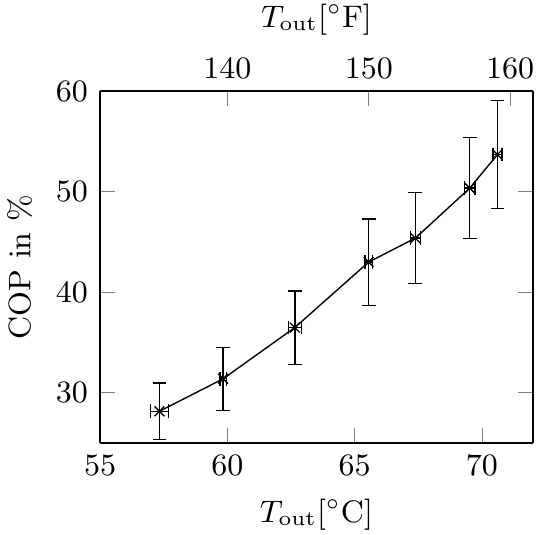}}\hfill
  \vspace*{-3mm}
  \caption{\subref{fig:increase} Relative node power increase for 13
    nodes with six-core E5645 processors and \subref{fig:cop} COP of
    adsorption chiller.  The vertical error bars in
    \subref{fig:increase} are the standard deviations after averaging
    over nodes, while the vertical error bars in \subref{fig:cop}
    reflect the 10\% accuracy of the flow meters.}
\end{figure}

Some of our measurements [those presented in Figs.~\ref{fig:core},
\ref{fig:power}, and \ref{fig:increase}] were taken on a subset of 13
randomly selected nodes (six-core E5645 processors at 2.4 GHz with
Turbo Boost disabled) running a well-defined load (the standard
\texttt{stress} tool \cite{stress}).  The other measurements were
taken on the whole iDataCool system running in production mode, i.e.,
various jobs of different sizes and with different computing and
communication requirements are scheduled and executed by the batch
queueing system.

In Fig.~\ref{fig:core} we show the average compute core temperature as
a function of the outlet temperature.  The average difference between
core and water temperature increases slightly, from
15$^\circ$C/59$^\circ$F to 17.5$^\circ$C/63.5$^\circ$F, over the range
of temperatures considered.  The error bars are rather large,
indicating a large variation between nodes.  A histogram of core
temperatures for $T_{\text{out}}=67^\circ$C/153$^\circ$F is shown in
Fig.~\ref{fig:coredist}.  The solid line is a Gaussian fit centered at
$84^\circ$C/183$^\circ$F with $\sigma=2.8^\circ$C/5.0$^\circ$F.  The
small bump at the low end of the histogram is due to idle nodes that
have a much lower core temperature.  Our interpretation of the large
spread visible in Fig.~\ref{fig:coredist} is that it is mainly due to
the first segment of the heat transfer path described in
Sect.~\ref{sec:idc-arch}, over which we have no control, while we can
control the second part very carefully.  Nevertheless, this spread is
a real problem if we aim, with energy reuse in mind, for a high outlet
temperature.  The cores throttle at about
100$^\circ$C/212$^\circ$F,\footnote{Note that there are other
  processors that throttle already at much lower temperatures.  Such
  processors are obviously not suitable for cooling with hot water.}
so the outlet temperature is limited by the core with the largest
difference between core and outlet temperature.  In our system this
largest difference is below 30$^\circ$C/54$^\circ$F so that we can
safely run at $T_{\text{out}}\le70^\circ$C/158$^\circ$F.  If we
desired higher temperatures we could sort out the ``bad'' chips and
run them at lower temperature in a separate system.  The high end of
the histogram in Fig.~\ref{fig:coredist} indicates that we could
perhaps gain another 5$^\circ$C/9$^\circ$F in this way.

The power consumption per node also shows large fluctuations.  In
Fig.~\ref{fig:power} we present the DC power consumption of 13 nodes
vs.{} their average core temperature.  To quantify the spread we
measure the DC power on most six-core nodes for various temperatures,
interpolate to $80^\circ$C/176$^\circ$F, and then construct a
histogram of the interpolated node power, see
Fig.~\ref{fig:powerdist}.  The solid line is a Gaussian fit centered
at 206W with $\sigma=5.4$W.  We see that the individual CPUs vary
greatly in their power consumption even for the same coolant
temperature.  We again attribute most of these variations to the
manufacturing process of the chips, not to our liquid-cooling
solution.

With higher cooling-water temperatures the power consumption of the
nodes increases, which has a negative effect on the total energy reuse
efficiency of the system.  To quantify this effect we plot in
Fig.~\ref{fig:increase} the relative average increase of the node
power consumption, which is about 7\% when going from
49$^\circ$C/120$^\circ$F to 70$^\circ$C/158$^\circ$F.  This should be
compared to the efficiency gain of the adsorption chiller, which is
quantified by the COP defined in Sect.~\ref{sec:installation} and
shown in Fig.~\ref{fig:cop}.  The temperature in the last plot starts
at 57$^\circ$C/135$^\circ$F since the adsorption chiller is in standby
mode for lower temperatures.  When going from 57$^\circ$C/135$^\circ$F
to 70$^\circ$C/158$^\circ$F the COP increases by 90\%, while the node
power consumption increases by only 5\%.  Thus the energy reuse
efficiency dramatically improves when running at higher temperatures.

\begin{figure}[t]
  \centering
  \subfigure[][\label{fig:heat}]%
  {\includegraphics{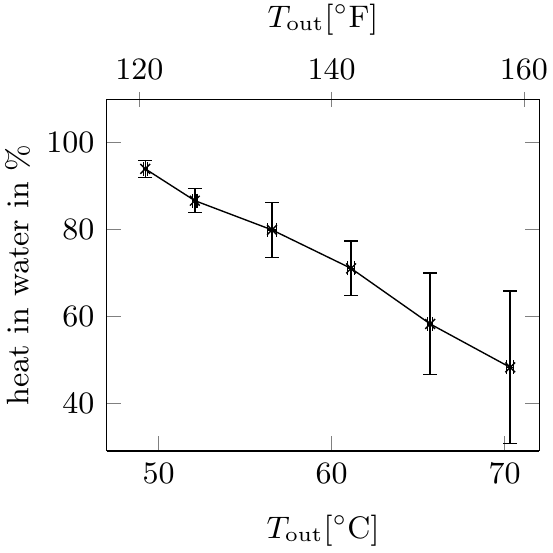}}\hfill
  \subfigure[][\label{fig:etrans}]%
  {\includegraphics{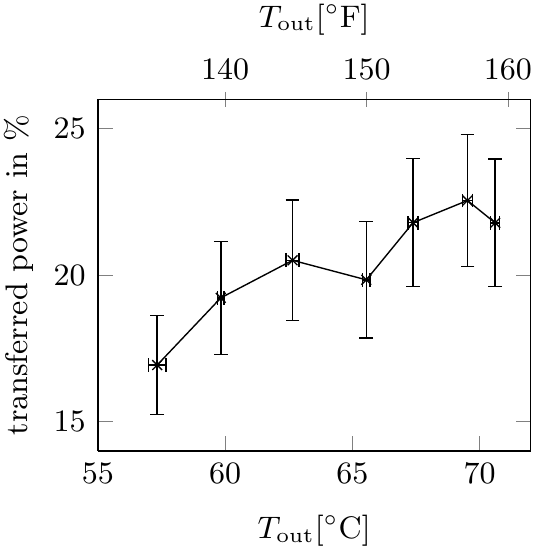}}
  \vspace*{-3mm}
  \caption{\subref{fig:heat} Heat-in-water fraction and
    \subref{fig:etrans} transferred power.  The vertical error bars in
    \subref{fig:heat} combine temporal fluctuations of the inlet- and
    outlet coolant temperatures and the flow, while the vertical error
    bars in \subref{fig:cop} reflect the 10\% accuracy of the flow
    meters.}
\end{figure}

In Fig.~\ref{fig:heat} we show the fraction of the electric power
delivered to the cluster that is transferred to the water in the rack
circuit.  We observe that this fraction drastically decreases with
temperature.  The reason for this decrease is the imperfect thermal
insulation of the iDataCool racks from the environment.\footnote{We
  did make serious insulation efforts, but since we retrofitted an
  existing system we were limited in what we could do.}  A higher
temperature difference between rack and air implies that more energy
is lost to the air.  The lesson from this figure is that in future
hot-water cooling designs serious attention should be paid to the
thermal insulation of the rack already in the early planning stages.
In Fig.~\ref{fig:etrans} we show the fraction of electric power that
is transferred to the driving circuit of the adsorption chiller, i.e.,
$P_d/P_\text{electric}$, as a function of the coolant temperature in
the rack circuit.\footnote{The energy balance in the rack circuit is
  $P_r=P_d+P_\text{add}+P_\text{loss}$, where 
  $P_r =$ heat-in-water $\times\; P_{\text{electric}}$,
  $P_\text{add}$ is the additional cooling power
  from the primary cooling circuit, and $P_\text{loss}$ is the heat
  per unit time that is lost to the environment due to imperfect
  thermal insulation of the plumbing.  $P_\text{add}$ is small at high
  temperatures, see Sect.~\ref{sec:installation}.  The fact that the
  numbers in Fig.~\ref{fig:etrans} are significantly lower than those
  in Fig.~\ref{fig:heat} thus implies that for our system
  $P_\text{loss}$ is rather large.} The increase shows that higher
coolant temperatures in the rack circuit lead to a better utilization
of the chiller, i.e., for our system the increase of the chiller
effectiveness with $T_{\text{out}}$ outweighs the reduced heat in
water.

We do not show plots of the fraction of energy reused (or,
equivalently, of the energy reuse efficiency) since the cooling
capacity of our chiller is not high enough to convert all heat
from the iDataCool system to chilled water.  The fraction of energy
that could be reused (e.g., by adding another chiller) can be computed
by multiplying the numbers in Figs.~\ref{fig:cop} and \ref{fig:heat}
and is on the order of 25\% for
$T=60\ldots70^\circ$C/140\ldots158$^\circ$F.  With better thermal
insulation this fraction could increase by almost a factor of two at
$T=70^\circ$C/158$^\circ$F, see Fig.~\ref{fig:heat}.


%% file: conclusions.tex
We have demonstrated that, by employing a sophisticated but low-cost
water-cooling solution, it is possible to cool a large compute cluster
in stable production mode with water outlet temperatures of up to
70$^\circ$C/158$^\circ$F.  At such temperatures a significant fraction
of the energy consumed by the cluster can be reused.  In the iDataCool
system the waste heat from the cluster drives an adsorption chiller
that operates efficiently above 65$^\circ$C/149$^\circ$F.  The minor
increase in power consumption of the nodes due to the higher
temperature is more than offset by the dramatic increase in the COP of
the chiller.

The main problem of the iDataCool system is the imperfect thermal
insulation, which leads to a serious loss of heat to the environment
and decreases the amount of energy that can be reused.  In future
designs this problem should be attacked from the very start.  Our
numbers indicate that with better thermal insulation almost 50\% of
the energy can be recovered in the form of chilled water.  Of course,
other opportunities for energy reuse exist where an even higher
fraction of the energy can be recovered, e.g., by heating.  However,
at some sites heating may not be an option or not necessary at all, in
which case the generation of chilled water, which can be used to cool
other parts of the computing center, is an attractive possibility.

Finally, an important issue is the effect of high water temperatures
on the reliability of electronic components, and in general on the
long-term stability of the system.  iDataCool gives us a unique
opportunity to study this issue (except for hard disks since the
iDataCool nodes are diskless).  We cannot predict the future, but
after more than one year of cooling with hot water we have not yet
observed any negative effects.


%% file: acknowledgment.tex
\sloppy

We thank J. Marschall (IBM Germany) and S.~Heybrock, B.~Mendl,
A.~Sch\"afer, and M.~Wimmer (University of Regensburg) for their
contributions to the project, and A. Auweter and H. Huber (LRZ
Garching) for helpful discussions. We also thank the machine shop
of the Regensburg Physics Department and InvenSor for technical
support.  The iDataCool project was funded by the German Research
Foundation (DFG), the German state of Bavaria, and IBM.

\fussy
